\documentclass[a4paper,12pt]{article}
\usepackage[utf8]{inputenc}
\usepackage[T1]{fontenc}
\usepackage{lmodern}
\usepackage{microtype}
\usepackage{cancel}
\usepackage[normalem]{ulem}
\usepackage{amsmath, amsthm, amsfonts}
\usepackage{amssymb}
\usepackage{graphicx}

\usepackage{enumerate}
\usepackage{mathtools}
\usepackage{subfig}
\usepackage{color}
\usepackage{float}
\usepackage{here}
\usepackage{cite}
\usepackage{mathrsfs}
\usepackage{epsfig}
\usepackage{dcolumn}
\usepackage{booktabs}

\usepackage{bm}
\usepackage[colorlinks=true,linkcolor=blue,citecolor=red]{hyperref}
\usepackage[toc,page]{appendix}
\newcommand{\be}{\begin{equation}}
\newcommand{\ee}{\end{equation}}
\newcommand{\bea}{\setlength\arraycolsep{2pt} \begin{eqnarray}}
\newcommand{\eea}{\end{eqnarray}}

\def\0{{\sst{(0)}}}
\def\1{{\sst{(1)}}}
\def\2{{\sst{(2)}}}
\def\3{{\sst{(3)}}}
\def\4{{\sst{(4)}}}
\def\5{{\sst{(5)}}}
\def\6{{\sst{(6)}}}
\def\7{{\sst{(7)}}}
\def\8{{\sst{(8)}}}
\def\sst#1{{\scriptscriptstyle #1}}

\renewcommand{\phi}{\varphi}

\makeatletter \@addtoreset{equation}{section}

\usepackage{multirow}
\newcommand{\ffuno}{\mbox{$J_1$}}
\newcommand{\ffdos}{\mbox{$J_2$}}
\newcommand{\dotff}{\mbox{$\dot{J}$}}

\newcommand{\epsilonzero}{\mbox{$\epsilon_1$}}
\newcommand{\epsilonuno}{\mbox{$\epsilon_2$}}
\newcommand{\deltazero}{\mbox{$\Delta_1$}}
\newcommand{\deltauno}{\mbox{$\Delta_2$}}
\newcommand{\kkzero}{\mbox{$k_1$}}
\newcommand{\kkuno}{\mbox{$k_2$}}

\begin{document}
%

\title{\normalsize
\phantom{fff}
\vspace{-3cm}
\begin{flushright}
{FISPAC-TH/31416/2026\\
UQBAR-TH/2026-27}
\end{flushright}
\vspace{2cm}
{\bf \Large Reconstructing slow-roll scalar-tensor Gauss-Bonnet single-field
inflation from running spectral data}}
\author{   \small
A. Belhaj$^{1}$\thanks{a-belhaj@um5r.ac.ma}, H. Es-Sobbahi$^{2}$\thanks{essobbahi@um5r.ac.ma},
M. Oualaid$^{1}$\thanks{oualaid@um5r.ac.ma},  E. Torrente-Lujan$^{3,4}$\thanks{torrente@cern.ch}\footnote{ Authors in alphabetical order.}
	\hspace*{-8pt} \\
	{\small $^1$ Dept. de Physique, Equipe des Sciences de la mati\`ere et du rayonnement,
		ESMaR}\\ {\small Faculty of Science, Mohammed V University, Rabat, Morocco}
\\ {\small $^2$ LHEP-MS, Faculty of Science, Mohammed V University, Rabat, Morocco} \\
 {\small $^{3}$ IFT, Dep. de F\'isica, Univ.  de Murcia,
Campus de Espinardo, E-30100 Murcia, Spain}
\\ {\small $^{4}$ TH-division, CERN, CH-1211 Geneva 23, Switzerland}
 }

\date{Revised after referee comments, May 17, 2026}
\maketitle
	\begin{abstract}
		{\noindent}
We study slow-roll inflation in a broad class of scalar-tensor theories containing scalar-dependent non-minimal derivative couplings to the Einstein tensor and scalar couplings to the four-dimensional Gauss-Bonnet invariant.  Working in the slow-roll regime, we derive theoretical expressions for the scalar and tensor spectral indices, the tensor-to-scalar ratio, the scalar-amplitude normalization, and the first and higher runnings in terms of the functions that define the model.  We also present consistency hierarchies relating scalar and tensor perturbations to the running parameters.  For a monomial toy model we obtain explicit formulae, discuss the range of validity of the slow-roll estimate near the end of inflation, and compare representative predictions with the final Planck 2018 constraints and the BICEP/Keck BK18 bound on primordial tensors.

	\end{abstract}

\newpage
\tableofcontents

\section{\label{intro}Introduction}

Einstein's theory of gravity, general relativity (GR), has passed a wide range of tests in both weak- and strong-field regimes, from late-time cosmology to compact-object systems.  Recent examples include the direct observation of gravitational waves \cite{Abbott:2016blz} and the Event Horizon Telescope observations of the M87 black-hole shadow, which are consistent with the Kerr black-hole picture in Einstein-Hilbert gravity \cite{Akiyama:2019cqa,Akiyama:2019eap}.  Much less is known about the behaviour of gravity at the very earliest stages of the Universe, especially during a possible inflationary era, where higher-curvature terms or non-minimal scalar interactions may become relevant.

The cosmic inflation scenario, beginning with the internally self-consistent quasi-de Sitter model of Starobinsky and followed by the Guth and Albrecht--Steinhardt models \cite{starobinski80,guth,steinhardt}, remains one of the most successful frameworks for the pre-Big Bang evolution of the Universe.  It provides a natural explanation of the flatness, horizon, and monopole problems of the standard hot Big Bang model and supplies the primordial fluctuations that seed large-scale structure and the observed cosmic microwave background (CMB) anisotropies \cite{revlinde,liddle,riotto,lyth,mukhanov,baumann}.  The calculation of inflationary perturbations has its own chronology: tensor perturbations were first computed quantitatively in Ref.~\cite{starobinsky3}, Bardeen's gauge-invariant formalism \cite{bardeen0} provided a key tool, and scalar perturbations were developed in Refs.~\cite{mukhanov1,hawking,starobinsky2,guth1,bardeen,starobinsky1}.  The near scale invariance of the scalar power spectrum has been confirmed with increasing precision by successive CMB data releases \cite{planck13,planck15,planck18}.

Given the observational CMB data, both bottom-up and top-down approaches are useful.  Bottom-up reconstructions attempt to infer the inflaton potential and possible deformations of Einstein gravity directly from the data, using Taylor expansions, splines, non-parametric methods, or effective Lagrangians with broad classes of allowed operators.  Top-down analyses, by contrast, test specific extensions of minimal GR with a small number of parameters.  Such toy models can isolate the phenomenological role of additional interactions and can clarify which classes of terms are still compatible with current data \cite{Fomin:2018blx,Hikmawan:2021hvc,Hikmawan:2019idy,Cortes:2006ap,Cortes:2007ak,Wolfson:2019rwd,Chen:2019eat}.

One of the robust lessons of the Planck data is that the scalar spectral index is smaller than unity while its running is consistent with zero.  The final Planck 2018 inflation analysis gives $n_S=0.9649\pm0.0042$ at 68\% confidence level and finds no evidence for a scale dependence of $n_S$ \cite{planck18}.  The scalar-amplitude normalization, $A_s\simeq 2.1\times 10^{-9}$ at the usual CMB pivot scale, provides an additional constraint on the energy scale of inflation \cite{planck18params}.  The Planck-only 95\% upper limit $r_{0.002}<0.10$ is strengthened when combined with BICEP/Keck data; the later BK18 analysis gives $r_{0.05}<0.036$ at 95\% confidence level \cite{BK18}.  These increasingly stringent bounds strongly constrain simple large-field monomial potentials and motivate the study of scalar-tensor scenarios in which non-minimal derivative and Gauss-Bonnet couplings can reduce the tensor amplitude.

The aim of this work is to investigate slow-roll scalar-tensor inflation from running spectral data.  In the slow-roll approximation, primordial perturbations can be related to the scalar potential, to the non-minimal scalar-dependent couplings, and to their derivatives.  This makes it possible to constrain the functions defining the model directly from observables.  More precisely, we derive expressions for $n_S$, $n_T$, the tensor-to-scalar ratio $r$, and the corresponding running quantities in scalar-tensor models that include non-minimal kinetic couplings to the Einstein tensor and couplings to the Gauss-Bonnet invariant.

The structure of the paper is as follows.  In Sec.~\ref{sec:2} we introduce the scalar-tensor Gauss-Bonnet model, derive the background equations in a spatially flat Friedmann--Robertson--Walker (FRW) geometry, and define the slow-roll hierarchy.  In Sec.~\ref{sec:3} we discuss scalar and tensor perturbations and present consistency relations among the spectral parameters and their runnings.  In Sec.~\ref{sec:4} we analyse a monomial toy model and compare representative numerical results with observational constraints.  Section~\ref{sec:5} contains the conclusions.

\section{The model and background equations}
\label{sec:2}
We investigate generic scalar-tensor models with a minimal Einstein-Hilbert sector, a non-minimal derivative coupling of the scalar field to the Einstein tensor, and a scalar coupling to the four-dimensional Gauss-Bonnet (GB) invariant.  The associated action, written in the Einstein frame, is
\begin{equation}
\label{e0021}
S=\int d^4x\sqrt{-g}\left[\frac{1}{2\kappa^2}R-\frac{1}{2}\partial_{\mu}\varphi\partial^{\mu}\varphi-V(\varphi)
+\ffuno(\varphi)G_{\mu\nu}\partial^{\mu}\varphi\partial^{\nu}\varphi-\ffdos(\varphi){\cal G}(R)\right]
\end{equation}
where $G_{\mu\nu}$ is the Einstein tensor and ${\cal G}$ is the four-dimensional GB invariant
\begin{equation}
{\cal G}=R^2-4R_{\mu\nu}R^{\mu\nu}+R_{\mu\nu\lambda\rho}R^{\mu\nu\lambda\rho},
\end{equation}
where we take $\kappa^2=M_p^{-2}=8\pi G=1$.  The field $\varphi$ is a real scalar with potential $V(\varphi)$; the differentiable functions $\ffuno(\varphi)$ and $\ffdos(\varphi)$ specify the non-minimal kinetic and Gauss-Bonnet sectors, respectively.  The symbol $R$ denotes the Ricci scalar.  A possible cosmological constant $\Lambda$ can be included in the scalar potential $V$.
The family of models defined by the action \eqref{e0021} depends on three scalar-dependent quantities, $V$, $\ffuno$, and $\ffdos$, which in a complete theory would be fixed by ultraviolet physics.  Here they are left arbitrary, subject only to the requirement that the resulting inflationary observables be compatible with the data \cite{planck13,planck15,planck18,BK18}.


In what follows, we assume a spatially flat FRW, homogeneous and isotropic, background  metric in any of the forms
\begin{eqnarray}
ds^2&=&-dt^2+a(t)^2 d\vec{r}^2 \label{e3001}
\end{eqnarray}
or
$ds^2= a(\eta)^2 (-d\eta^2+ d\vec{r}^2)$
where $a$ is a scale factor. 
The metric and  the scalar field equations of motion can be written as follows , using \eqref{e3001},\cite{Fomin:2018blx,grandajimenez,Chen:2019eat}
\begin{eqnarray}
\label{e0025}
3{H^2}&=&  \frac{1}{2}{{\dot \varphi }^2} + V  +9H^2 \ffuno\dot{\varphi}^2+24H^3\dotff_2,
\\
\label{e0026}
-2\dot{H}&=&\dot{\varphi}^2+
6H^2\ffuno\dot{\varphi}^2-2\frac{d}{dt}\left(H\ffuno\dot{\varphi}^2\right)+
8H^3 \dotff_2 -8 \frac{d}{dt}(H^2\dotff_2),
\\
\label{e0027}
\ddot \varphi  + 3H\dot \varphi  + V'&=&
- 24{H^2}\left( {{H^2} + \dot H} \right)\ffdos'
- 6H{\ffuno}\dot \varphi (3{H^2} + 2\dot H)\nonumber\\ &&
- 3{H^2}\left( {2{\ffuno}\ddot \varphi  + \ffuno'{{\dot \varphi }^2}} \right)
\end{eqnarray}
where the dot represents derivatives with respect to the time variable $t$,
$H\equiv \dot{a}/a$ is the Hubble parameter and  $\ffuno'$ and $\ffdos'$ are derivatives with respect $\varphi$.
Only two of the previous equations are independent due to the existence of  the Bianchi identities.
Taking  $\ffuno=\ffdos=0$,  we recover explicitly the standard Friedmann equations with a single scalar field
\begin{eqnarray}
3{H^2}&=&\frac{1}{2}{{\dot \varphi }^2} + V,\\
-2\dot{H}&=&\dot{\varphi}^2,\\
\ddot \varphi  + 3H\dot \varphi  + V' &=& 0.
\end{eqnarray}
After a simple computation,  we get from the   field equations \eqref{e0025}-\eqref{e0027} expressions 
 for $\dot{\varphi}^2$ and $V$ as  follows
\begin{eqnarray}
V&=&H^2\left(3-\epsilonzero-\frac{5}{2}\deltazero-2\kkzero-\frac{1}{2}\deltazero\left(\deltauno-\epsilonzero\right)-\frac{1}{3}\kkzero\left(\kkuno-\epsilonzero\right)\right),\\
\dot{\varphi}^2&=&H^2\left(2\epsilonzero-\deltazero-2\kkzero+\deltazero\left(\deltauno-\epsilonzero\right)+\frac{2}{3}\kkzero\left(\kkuno-\epsilonzero\right)\right)
\end{eqnarray}
where we  have used the parameter definitions\footnote{ We use the conventions of the PLANCK collaboration \protect\cite{planck18}.}
\begin{eqnarray}
\epsilonzero&=&\dot{\left(\frac{1}{H}\right)}=-\frac{\dot{H}}{H^2},\quad\\
\kkzero&=&3 \ffuno\dot{\varphi}^2,  \quad\\
\deltazero&=&8H\dot{\ffdos}.
\end{eqnarray}
For further use, we define also the following  parameter hierarchy: for any quantity $X$, in particular for
$X=\epsilonzero,\kkzero,\deltazero$, we define ($n\geq 1$) using the number of e-folds remaining until the end of inflation, $N\equiv \ln(a_e/a)$, so that $dN=-Hdt$
\begin{equation}
X_{n+1} = -\frac{d \log \mid X_n\mid }{dN}=\frac{d \log \mid X_n\mid }{d\log a}.
\end{equation}
This set of parameters satisfies the following  properties, for any $X_n$
\begin{eqnarray}
\label{Eq 2.14}
-\frac{d  X_n}{dN} &=& X_n X_{n+1}=-X_n' \varphi_{,N},\\
\label{Eq 2.15}\frac{d^2 X_n}{dN^2} &=& X_n X_{n+1}\left(X_{n+1}+X_{n+2}\right).
\end{eqnarray}
where $X_{,N}\equiv dX/dN$. The main reason for the use of these definitions is the following. In terms of the wavenumber $k$,
 defined by $ k\equiv a H$, we have
\begin{eqnarray}
\frac{d X}{d N} = (\epsilonzero-1) \frac{d X}{d \log k}.
\end{eqnarray}
Differentiating once more gives
\begin{eqnarray}
\frac{d^2 X}{d N^2} = -\epsilonzero\epsilonuno \frac{d X}{d \log k}
+(1-\epsilonzero)^2 \frac{d^2 X}{d(\log k)^2}.
\end{eqnarray}
Equivalently,
\begin{eqnarray}
\frac{d X}{d \log k} &=& \frac{1}{(\epsilonzero-1)}\frac{d X}{d N},\\
\frac{d^2 X}{d(\log k)^2} &=&
\frac{1}{(\epsilonzero-1)^2}\left(\frac{d^2 X}{d N^2}
+\frac{\epsilonzero\epsilonuno}{(\epsilonzero-1)}\frac{d X}{d N}\right).
\end{eqnarray}
For small values of the slow-roll parameters ($\epsilonzero\to 0$), these relations reduce to
\begin{eqnarray}
\frac{d X}{d \log k} &\simeq&-\frac{d X}{d N},\\
\frac{d^2 X}{d(\log k)^2} &\simeq& \frac{d^2 X}{d N^2}.
\end{eqnarray}
Near the end of inflation, when $\epsilonzero\to 1$, the first derivative with respect to $N$ is suppressed.

A quasi-exponential inflationary expansion is provided, in particular,  in the slow-roll regime. This regime is defined by the 
conditions
 $\epsilon_i, k_i, \Delta_i\ll 1$ (slow-roll conditions). Under this assumption,
 the field equations \eqref{e0025}-\eqref{e0027} are greatly simplified. They become
\begin{eqnarray}
3H^2- V &\simeq & 0,\\
\label{Eq 2.17}\dot{H}+\frac{1}{2}\dot{\varphi}^2&\simeq &-3H^2\ffuno\dot{\varphi}^2-4H^3\dotff_2 =-\frac{H^2}{2}( 2 \kkzero+\deltazero),\\
\label{Eq 2.18}\dot{\varphi}+\frac{V'}{3H}&\simeq &-6H^2\ffuno\dot{\varphi}-8 H^3 \ffdos'.
\end{eqnarray}


 
At the slow-roll level there is a degeneracy among models: different triplets $(V,\ffuno,\ffdos)$ in the action \eqref{e0021} can lead to the same observational results.  To isolate the effective dependence on the scalar-dependent functions, it is convenient to rewrite the slow-roll field equations \eqref{Eq 2.17} and \eqref{Eq 2.18} using the number of e-folds $N$ as the independent variable:
\begin{eqnarray}
\epsilonzero &=& \frac{\dot\varphi^2}{2 H^2} (1+6 H^2 \ffuno)+ 4 H \dot \ffdos,\\
 -H\frac{d\varphi}{dN}+\frac{V'}{3H}&\simeq &-6H^2\ffuno\dot{\varphi}-8 H^3\ffdos'
\end{eqnarray}
which can be written as
\begin{eqnarray}
\epsilonzero &=& \frac{1}{2} \left (\frac{d \varphi}{dN}\right )^2\left ( 1+2 V \ffuno\right )-
\frac{4}{3} V\ffdos' \frac{d \varphi}{dN},\\
\frac{d\varphi}{dN}&=&\frac{ V'}{V}\frac{ 1+8 \ffdos' V^{2}/3V'}{1+2 V\ffuno}.
\end{eqnarray}
It follows from such  expressions that the model  depends only on the scalar combinations
\begin{equation}v\equiv V'/V,\quad f_1\equiv V \ffuno,\quad f_2\equiv V\ffdos'.
\end{equation}
 In terms of the reduced variables $v,f_1,f_2$ and the quantity
$$f_3\equiv \frac{f_2}{v},$$
 we can write
\begin{eqnarray}
\epsilonzero &=& \frac{1}{2} \left (\frac{d \varphi}{dN}\right )^2\left ( 1+2 f_1\right )-\frac{4}{3} f_2\frac{d \varphi}{dN},\\
\frac{d\varphi}{dN} &=& v\,\frac{1+\frac{8}{3}f_3}{1+2 f_1}.
\end{eqnarray}
The number of e-folds  reads as
\begin{equation}
N=\int_{\varphi_i}^{\varphi_e}
\frac{1}{v}\frac{1+2 f_1}{ 1+\frac{8}{3}f_3}d\varphi
\end{equation}
where $\varphi_i$ and $\varphi_e$ are the values of the scalar field at the beginning and  at the  end of the inflation period,  respectively.
The rest of slow-roll parameters can also be written in terms of $v,f_1,f_2$ and $f_3$. 
After a straightforward computation,  we find  the following relevant  expressions
\begin{eqnarray}
\epsilonzero &=& \frac{v}{2}  \varphi_{,N}= \frac{v^2}{2}\frac{ (1+\frac{8}{3}f_3)}{1+2 f_1},\\
\kkzero &=&   f_1 \, \left (\varphi_{,N}\right)^2=      f_1 \,v ^2 \, \frac{ \left(1+\frac{8}{3}f_3\right){}^2 }{\left(1+2 f_1\right){}^2},\\
\deltazero &=& -\frac{8}{3} f_2 \, \varphi_{,N}=       -\frac{8 f_2 \, v}{3}\frac{ \left(1+\frac{8}{3}f_3\right)  }{ \left(1+2 f_1\right)}.
\end{eqnarray}
Higher order parameters, for example $\epsilonuno,\kkuno,\deltauno$, are easily obtained using Eq (\ref{Eq 2.14}).
Successful inflation occurs when $0<\epsilonzero< 1$.
The limit $\epsilonzero(\varphi_i)\to0$ depends strongly on the Gauss-Bonnet-related combination $f_3=f_2/v$.  Small values of $\epsilonzero(\varphi_i)$ can be reached in two representative regimes.  First, if
\begin{eqnarray}
\varphi_{,N}\simeq 0,
\end{eqnarray}
then the Gauss-Bonnet-related combination satisfies
  \begin{eqnarray}   f_3\simeq -3/8 .\end{eqnarray}
Second, when
\begin{eqnarray} f_3=-1/2,\end{eqnarray}
one requires
 \begin{eqnarray}
  \varphi_{,N}\simeq -\frac{v}{3} \frac{1}{1+2 f_1}.
  \end{eqnarray}
The graceful exit of inflation depends on whether the equation $\epsilonzero(\varphi_e)=1$
can be accomplished for some $\varphi_e$. The graceful exit relation becomes
\begin{equation}
\label{Eq 2.30}
\epsilonzero(\varphi_e)=1= \frac{v^2}{2} \frac{ (1+\frac{8}{3}f_3)}{1+2 f_1}\Big |{}_{\varphi_e}.
\end{equation}
Using $f_3=f_2/v$, this relation can be written as a quadratic equation for $v$,
\begin{equation}
\frac{1}{2}v^2+\frac{4}{3}f_2 v-(1+2f_1)=0,
\end{equation}
whose real solutions require
\begin{equation}
(f_2)^2\geq -\frac{9}{8}\left(1+2 f_1\right)
\end{equation}
at $\varphi=\varphi_e$.  In terms of the functions in the action this condition reads
\begin{equation}
(V\ffdos')^2 \geq -\frac{9}{8}\left(1+2V\ffuno\right).
\end{equation}
In particular, $V(\varphi_e)\ffuno(\varphi_e)\geq -1/2$ is a simple sufficient condition for the reality of the exit solution.

\section{First- and higher-order perturbations and running consistency equations}
\label{sec:3}

The spectral indices are defined by \cite{planck18,Fomin:2018blx,Cortes:2006ap,Cortes:2007ak}
\begin{eqnarray}
n_S-1&=&\frac{d\log P_S}{ d\log k},\\
n_T&=&\frac{d\log P_T}{ d\log k}.
\end{eqnarray}
where $P_S$ and $P_T$ are the power spectra of the scalar and tensor modes.
The running and the running of the running of the spectral indices are
given by the following expressions in terms of the slow-roll parameters   
\cite{Cortes:2006ap,Cortes:2007ak,grandajimenez}
\begin{eqnarray}
n_S-1 &=& -2\epsilonzero-\frac{2\epsilonzero\epsilonuno-\deltazero\deltauno}{2\epsilonzero-\deltazero},\label{e0033}\\
\label{e0034}
n_T &=& -2\epsilonzero.
\end{eqnarray}
Moreover,  the tensor-to-scalar ratio  takes  the following form \cite{grandajimenez}
\begin{equation}
\label{e0035}
r=8\left(\frac{2\epsilonzero-\deltazero}{1-\frac{1}{3}\kkzero-\deltazero}\right)
\simeq 8\left(1-\frac{1}{3}\kkzero-\deltazero\right)\left(2\epsilonzero-\deltazero\right)\simeq 8\left(2\epsilonzero-\deltazero\right)+O^2,
\end{equation}
where $O^2$ are terms quadratic in the slow-roll parameters.
In this class of models, as in standard GR single scalar inflation, the parameters
 $n_S$, $n_T$ and $r$ are not independent variables. The models   predict a relation among them which reads as
\begin{equation}
r+8n_T \simeq -8\deltazero+O^2.
\end{equation}
This relation differs
 from the standard inflation relation $r=-8 n_T$   by $\deltazero$, 
 this is a   supposedly small quantity,
proportional to the GB term
\begin{equation}
r+8n_T \simeq -64 H \dot{J_2}.
\end{equation}

The amplitude of scalar perturbations is an additional observable normalization, not only a derived quantity.  To the same leading slow-roll order used in this work, we define
\begin{equation}
A_s\equiv P_S(k_*)\simeq \frac{H_*^2}{4\pi^2\left(2\epsilon_{1*}-\Delta_{1*}\right)}=\frac{H_*^2}{8\pi^2\epsilon_{\rm eff,*}},
\qquad \epsilon_{\rm eff,*}\equiv \epsilon_{1*}-\frac{1}{2}\Delta_{1*},
\label{eq:As}
\end{equation}
where the star denotes evaluation at horizon crossing for the pivot scale.  Combining Eq.~\eqref{eq:As} with the leading expression for $r$ gives
\begin{equation}
H_* \simeq \pi\sqrt{\frac{A_s r_*}{2}},\qquad
V_*\simeq 3H_*^2 \simeq \frac{3}{2}\pi^2 A_s r_* .
\label{eq:Hstar}
\end{equation}
Thus the observed value of $A_s$ fixes the overall scale of the potential once a point $(n_S,r)$ in the parameter scan has been selected.  For the monomial potential $V=\lambda\varphi^n/n$, this implies
\begin{equation}
\lambda \simeq \frac{3nH_*^2}{\varphi_*^n}=\frac{3n\pi^2A_s r_*}{2\varphi_*^n}.
\label{eq:lambdaAs}
\end{equation}
Corrections involving the scalar sound speed and the full Horndeski normalization are beyond the leading slow-roll truncation used here; Eq.~\eqref{eq:As} is the normalization appropriate to the same approximation as Eqs.~\eqref{e0033}--\eqref{e0035}.  Similar questions about the scalar-field evolution and the Hubble scale in Einstein--Gauss--Bonnet inflation have been discussed, for example, in Ref.~\cite{Oikonomou:2021kql}.

 It turns out that  we  can express observable quantities  like the
scalar and tensor power spectrum,  the spectrum index and tensor-to-scalar ratio in terms of
$v,f_1,f_2$ (evaluated at $\varphi=\varphi_i$). For    instance, 
the tensor spectrum index  and the tensor-to-scalar ratio are given  by 
\begin{eqnarray}
n_T &=&       - v^2\frac{ \left(1+\frac{8f_2}{3v}\right)}{1+2 f_1},\\ \label{e2002}
r &=& 8 v^2\frac{ \left(1+\frac{8f_2}{3v}\right)^2}{(1+2 f_1)}. \label{e2003}
\end{eqnarray}
meanwhile the scalar spectral index is written as
\begin{eqnarray}
n_{S}-1=-\frac{9\, v \, \left(I_{1}+v ^2\right)+24\, f_2\, \left(I_{2}+2 \,v ^2\right)+64 \,f_2^2 \,v +2 \,f_1 \, v \, \left(3 \, v +8\, f_2 \right){}^2}{3\, \left(1+2\, f_1\right){}^2 \left(3\, v + 8\, f_2 \right)} \label{e2001}
\end{eqnarray}
with 
\begin{eqnarray}
I_{1} &\equiv&      -\left(1+\frac{8 \,f_2}{3 \,v }\right) \left(\left(2+4 f_1\right)\, v '-2\, v f_1'\right)-\frac{1}{3} 8 \left(1+2 f_1\right)\, v  \left(\frac{f_2}{v }\right)'  ,\\
I_{2} &\equiv&       \frac{2}{3} \,f_2 \,\left(3+8 f_2\, v \right) f_1'-\left(1+2 f_1\right) \left(\left(1+\frac{8\, f_2}{3\,v}\right) \left(f_2\, v '+v  f_2'\right)+\frac{8}{3} f_2\, v \left(\frac{f_2}{v }\right)'\right).
\end{eqnarray}
In concrete models we can invert the relations \eqref{e2002}-\eqref{e2001} to get  the triplet of theoretical values $v,f_1,f_2$ in terms of the observable triplet
$(n_S,n_T,r)$,
or, in terms of the original variables $(V,\ffuno,\ffdos)$.

One can get further relations between the different spectral parameters and their runnings. 
For example, the physical scalar spectral index can be related to the scalar ratio, the running of the logarithm of the ratio and the
parameter $n_T$. 
It is straightforward to check that 
the equation \eqref{e2001} can be written as
\begin{eqnarray}
n_S-1 &=& -2\epsilonzero+\frac{d}{dN} \log(2\epsilonzero-\deltazero),\label{e00312}
\end{eqnarray}
This relation is the direct analogue, in the present scalar-tensor model with derivative coupling, of reconstruction formulae used in Einstein--Gauss--Bonnet inflation; compare, for instance, Eq.~(21) of Ref.~\cite{Pozdeeva:2020apf}.  In terms of a reconstruction function depending on $\varphi(N)$, one may equivalently rewrite the logarithmic term in Eq.~\eqref{e00312} as a derivative of the effective combination $2\epsilon_1-\Delta_1$, but in the present model that combination contains both $J_1$ and $J_2$ and is less compact than in pure Gauss--Bonnet models.
Using the expressions \eqref{e0035}, can be written as
\begin{eqnarray}
n_S-1= n_T+ (\epsilonzero-1) \frac{d \log r}{d \log k}+ O^2
\label{e00313}
\end{eqnarray}
where $O^2$ denotes higher-order slow-roll terms; here $O^2=\frac{1}{3}\kkuno+\deltauno$.
From \eqref{e00313}, when $\epsilonzero\simeq 1$, we obtain the constraint
\begin{eqnarray}
n_S-1= n_T+ O^2
\end{eqnarray}
or
\begin{eqnarray}
n_S-1= -r/8-\deltazero+\deltauno+\frac{1}{3} \kkuno.
\end{eqnarray}

Hierarchies of consistency equations relating scalar and tensor perturbations 
and low- and higher-order running parameters have been 
presented in Refs.~\cite{Cortes:2006ap,Cortes:2007ak}.
For the class of models considered here, the running and the running of the running of the scalar spectral index follow from \eqref{e00312} and \eqref{e00313}:
\begin{eqnarray}
 \frac{d n_S}{d \log a} &=& -\frac{1}{8} \frac{d r}{d\log a} - \frac{d^2\log r}{d(\log a)^2}
- \frac{d\deltazero}{d \log a},\\ \nonumber
&=& -\frac{r}{8} \left (\frac{d\log r}{d\log a}\right )  - \frac{d^2\log r}{d(\log a)^2}
- \frac{d\deltazero}{d \log a},\\
\frac{d^2 n_S}{d(\log a)^2}&=&-\frac{r}{8}\left[\left(\frac{d\log r}{d\log a}\right)^2+\frac{d^2\log r}{d(\log a)^2}\right]
- \frac{d^3\log r}{d(\log a)^3}- \frac{d^2\deltazero}{d(\log a)^2}.
\end{eqnarray}
Using Eqs.~(\ref{Eq 2.14}) and (\ref{Eq 2.15}), we obtain
\begin{eqnarray}
\frac{d\deltazero}{d \log a} &=& \deltazero \deltauno,\\
\frac{d^2\deltazero}{d(\log a)^2}&=& \deltazero \deltauno(\deltauno+\Delta_3).
\end{eqnarray}
In terms of the derivatives with respect to  the field $\varphi$, they become as follows
\begin{eqnarray}
 \frac{d \log r }{d \log a} &=& -\left (\log r\right)' \varphi_{,N},\\
 \frac{d^2\log r }{d(\log a)^2} &=& \left (\log r\right)'' (\varphi_{,N})^2+
\frac{1}{2}\left (\log r\right)' \left[ \left (\varphi_{,N}\right )^2\right]',
\end{eqnarray}
together with
\begin{equation}
\frac{d^3\log r }{d(\log a)^3} = -\left (\log r\right)''' (\varphi_{,N})^3 -\left( \frac{3}{2}\left (\log r\right)'' \left ((\varphi_{,N})^2\right)' + \frac{1}{2}\left (\log r\right)' \left ((\varphi_{,N})^2\right )''\right)   \varphi_{,N}.
\end{equation}

\section{A simple toy model: the monomial model}
\label{sec:4}

We now provide an explicit model with power-law potentials and scalar couplings.  In this case the expressions can be evaluated further.  We focus on monomial couplings parameterized as
\begin{equation}
V = \frac{\lambda}{n} \varphi^n,\quad \ffuno= \beta'_1\varphi^{m_1-n},\quad \ffdos= {\beta'_2} \varphi^{m_2-n}.
\end{equation}
where $\lambda$ is a coupling parameter.  The physical results depend only on the combinations
($\beta_1=\frac{\lambda\beta_1'}{n},\beta_2=\frac{\lambda\beta_2'}{n(m_2-n)} \ (m_2\not = n)$, $\beta_2=0$ for $m_2=n$ ) which are given by
\begin{equation}
v =\frac{n}{\varphi},\quad f_1=\beta_1 \varphi^{m_1},
\quad f_2=\beta_2 \varphi^{m_2-1}.
\end{equation}
Thus the slow-roll predictions effectively depend on the two real parameters $\beta_{1,2}$ and on the integer $n$.
Taking $c=\frac{8 \beta_2}{3n}$ and $b= 2\beta_1$, we get slow-roll parameters
\begin{eqnarray}
\epsilonzero &=& \frac{n^2}{2 \varphi^{2}}\frac{ (1+c \varphi^{m_2} )}{(1+b\varphi^{m_1})},\\  \kkzero
&=& \frac{n^2 b \varphi^{m_{1}}}{2 \varphi^{2}}\frac{ (1+c \varphi^{m_2} )^{2}}{(1+b\varphi^{m_1})^{2}},\\
\deltazero &=& \frac{-c n^{2}\varphi^{m_{2}}}{\varphi^{2}}\frac{ (1+c \varphi^{m_2} )}{(1+b\varphi^{m_1})}.
\end{eqnarray}
From Eq.~(\ref{Eq 2.14}), the next slow-roll parameters are
\begin{eqnarray}
\epsilon_{2}&=&\frac{n(1+c\varphi^{m_2})}{\varphi^{2}(1+b\varphi^{m_1})}
\left[2-\frac{c m_2\varphi^{m_2}}{1+c\varphi^{m_2}}+\frac{b m_1\varphi^{m_1}}{1+b\varphi^{m_1}}\right],\\
k_{2}&=&\frac{n(1+c\varphi^{m_2})}{\varphi^{2}(1+b\varphi^{m_1})}
\left[2-m_1-\frac{2c m_2\varphi^{m_2}}{1+c\varphi^{m_2}}+\frac{2b m_1\varphi^{m_1}}{1+b\varphi^{m_1}}\right],\\
\Delta_{2}&=&\frac{n(1+c\varphi^{m_2})}{\varphi^{2}(1+b\varphi^{m_1})}
\left[2-m_2-\frac{c m_2\varphi^{m_2}}{1+c\varphi^{m_2}}+\frac{b m_1\varphi^{m_1}}{1+b\varphi^{m_1}}\right].
\end{eqnarray}
In this way, the  number of e-foldings is
\begin{equation}
N =\int_{\varphi_{end}}^{\varphi_*}\frac{\varphi}{n} \frac{ 1+b \varphi^{m_1} }{1+c \varphi^{m_2}}d\varphi.
\end{equation}
The last expression can be integrated in a closed form  in order  to provide
\begin{eqnarray}
N-N_0 &=& \frac{\varphi^2}{2n} {}_2 F_1\left(1, \frac{2}{m_2},1+\frac{2}{m_2}, -c\varphi^{m_2}\right)\\ \nonumber
&&+\frac{b \varphi^{m_1+2}}{n(2+m_1)} {}_2 F_1
 \left(1, \frac{m_1+2}{m_2},1+\frac{m_1+2}{m_2}, -c \varphi^{m_2}\right)\\
&=&  \frac{\varphi^2}{2n} g_1(c,\varphi)+  \frac{b}{n}\frac{ \varphi^{m_1+2}}{2+m_1} g_2(c,\varphi) \nonumber
\end{eqnarray}
where ${}_2 F_1$ is the Hypergeometric function.
For small Gauss-Bonnet coupling ($c\to 0$), the functions $ g_1$ and $g_2$ reduce to
\begin{eqnarray}
g_1 &\simeq & 1-\frac{2c \varphi^{m_2}}{2+m_2}+ O(c)^2,\\
g_2&\simeq & 1-\frac{(2+m_1) c \varphi^{m_2}}{ 2+m_1+m_2}+ O(c)^2.
\end{eqnarray}
At the leading order in $c$ and $b$,  we find
\begin{equation}
N-N_0 \simeq\frac{\varphi^2}{2n} -c\frac{\varphi^{m_2+2}}{n(2+m_2)}+  \frac{b}{n} \frac{ \varphi^{m_1+2}}{2+m_1}.
\end{equation}
In the large coupling  ($c, b\to \infty$), however, we get
\begin{equation}
N-N_0 \simeq  \frac{b/c}{n (m_1-m_2+2)} \varphi^{m_1-m_2+2}.
\end{equation}
We now study a special subcase.  When $m_2=0$, the Gauss-Bonnet coupling takes the form $\ffdos\propto\varphi^{-n}$ for the monomial potential $V\propto\varphi^n$, and therefore $VJ_2$ is constant.  Such relations between the potential and the Gauss-Bonnet coupling have been considered in Gauss-Bonnet inflation starting with Ref.~\cite{Guo:2010jr}.  The non-minimal kinetic coupling remains essential in the present analysis: setting $\beta_1=0$ (or $b=0$) reduces the model to the minimally kinetic-coupled Gauss-Bonnet case, whereas $b\ne0$ suppresses $n_T$ and $r$ through the denominator $1+b\varphi^{m_1}$ and shifts $n_S$ through the derivative term proportional to $m_1$.  In this subcase the slow-roll parameters are
\begin{eqnarray}
\epsilonzero &=& \frac{n^2}{2\varphi^{2}}\frac{ (1+c)}{(1+b\varphi^{m_1})},\label{ep0}\\
\kkzero &=&\frac{n^2b\varphi^{m_{1}}}{2\varphi^{2}}\frac{ (1+c)^2}{(1+b\varphi^{m_1})^{2}},\\
\deltazero &=&\frac{-c n^2}{ \varphi^{2}}\frac{ (1+c)}{(1+b\varphi^{m_1})}.
\end{eqnarray}
For the next slow-roll parameters one obtains
\begin{eqnarray}
\epsilon_{2}&=&\frac{n(1+c)}{\varphi^{2}(1+b\varphi^{m_{1}})^2}
\left[2(1+b\varphi^{m_{1}})+b m_1\varphi^{m_{1}}\right],\\
k_{2}&=&\frac{n(1+c)}{\varphi^{2}(1+b\varphi^{m_{1}})^2}
\left[(2-m_1)(1+b\varphi^{m_{1}})+2b m_1\varphi^{m_{1}}\right],\\
\Delta_{2}&=&\frac{n(1+c)}{\varphi^{2}(1+b\varphi^{m_{1}})^2}
\left[2(1+b\varphi^{m_{1}})+b m_1\varphi^{m_{1}}\right].
\end{eqnarray}
For this model, the number of e-foldings takes the following form
\begin{eqnarray}
N &=& \int_{\varphi_{end}}^{\varphi_*} \frac{\varphi}{n} \frac{ 1+b \varphi^{m_1} }{1+c}d\varphi.
\end{eqnarray}
Integrating this expression, we find
\begin{eqnarray}
N&=& \frac{1}{n(1+c)}\left( \frac{\varphi^{2}}{2}+\frac{b \varphi^{m_{1}+2}}{m_{1}+2}\right)_{\varphi_{end}}^{\varphi_*}. \label{Nelf}
\end{eqnarray}

At leading slow-roll order, the scalar spectral index, the tensor spectral index, and the tensor-to-scalar ratio are
\begin{eqnarray}
n_{S}&=&1-\frac{n}{\varphi^{2}} \frac{b \, m_1 \varphi^{m_{1}}(1+c)
+(1+b \, \varphi^{m_{1}})(n \,(1+c)+2 \,c+2)}{(1+b \,\varphi^{m_{1}})^{2}}, \\
n_T &=&        - \frac{n^2}{\varphi^{2}}\frac{ (1+c)}{(1+b\,\varphi^{m_1})},\\
r&=&8\frac{n^{2}}{\varphi^2}\frac{(1+c)^2}{(1+b \, \varphi^{m_{1}})}.\label{e4r}
\end{eqnarray}


The running of the scalar spectral index is given by
\begin{eqnarray}
\frac{d n_S}{d \log k}=\frac{1}{(1-\epsilonzero)} \frac{d n_S}{d \log a},
\end{eqnarray}
with
\begin{eqnarray}
 \frac{dn_S}{ d\log a} &=&- n_S' \varphi_{,N},\\
\nonumber \frac{dn_S}{ d\log a}  &=&-\frac{n^2(1+c)^2 (A_1+A_2)}{A_3}.
\end{eqnarray}
where the terms $A_i$ are
\begin{eqnarray}
\nonumber A_1&=&(1+b \varphi ^{m_1})\Big((4+2n)(1+b \varphi ^{m_1})+ b m_1 \varphi ^{m_1}(4+n)\Big) \\
\nonumber  A_2&=& b m_1^2  \varphi^{m_1}(-1+b \varphi ^{m_1})\\
\nonumber  A_3&=& \varphi ^4 (1+b \varphi ^{m_1})^4.
\end{eqnarray}

The field value at the end of inflation is obtained by solving $\epsilon_1(\varphi_e)=1$ in Eq.~\eqref{ep0}, while $\varphi_*$ is then obtained from Eq.~\eqref{Nelf} for the chosen number of e-folds.  For the parameter scan used in the figures below, namely
\begin{equation}
 n=2,\qquad b=2.5\times 10^{-4},\qquad c=-0.92,\qquad 7\leq m_1\leq 10,
\label{eq:numericalscan}
\end{equation}
one obtains $\varphi_e\simeq0.400$ for the full displayed interval and $\varphi_*\simeq2.77$--$3.63$ for $N_*=50$--$60$.  Thus the scalar field rolls from $\varphi_*$ to $\varphi_e$ and does not tend to zero during the slow-roll portion displayed here.  It may subsequently approach the minimum at $\varphi=0$ only after the slow-roll regime, depending on the post-inflationary reheating dynamics, which is outside the scope of the present approximation.

The amplitude $A_s$ fixes the energy scale through Eq.~\eqref{eq:Hstar}.  Taking $\ln(10^{10}A_s)=3.044$, the representative points listed in Table~\ref{tab:numerics} give $H_*\simeq5.3\times10^{-6}$--$8.3\times10^{-6}$ and $H_e\simeq7.5\times10^{-7}$--$9.6\times10^{-7}$ in reduced Planck units.  At horizon crossing the slow-roll parameters in the same scan remain small:
\begin{equation}
0.0021\lesssim\epsilon_{1*}\lesssim0.0053,\qquad
1.5\times10^{-4}\lesssim k_{1*}\lesssim2.5\times10^{-4},\qquad
0.0039\lesssim\Delta_{1*}\lesssim0.0097,
\end{equation}
with $0.023\lesssim\epsilon_{2*}=\Delta_{2*}\lesssim0.033$ and $|k_{2*}|\lesssim0.026$.

\begin{table}[H]
\centering
\caption{Representative horizon-exit and end-of-inflation quantities for the scan in Eq.~\eqref{eq:numericalscan}.  The Hubble parameters are given in reduced Planck units and are obtained from the scalar-amplitude normalization.}
\label{tab:numerics}
\begin{tabular}{cccccccc}
\toprule
$m_1$ & $N_*$ & $\varphi_*$ & $\varphi_e$ & $H_*$ & $H_e$ & $n_S$ & $r$ \\ 
\midrule
7  & 50 & 3.474 & 0.400 & $8.34\times10^{-6}$ & $9.61\times10^{-7}$ & 0.9568 & 0.0067 \\ 
7  & 55 & 3.559 & 0.400 & $7.72\times10^{-6}$ & $8.68\times10^{-7}$ & 0.9617 & 0.0058 \\ 
7  & 60 & 3.635 & 0.400 & $7.20\times10^{-6}$ & $7.93\times10^{-7}$ & 0.9658 & 0.0050 \\ 
10 & 50 & 2.770 & 0.400 & $6.02\times10^{-6}$ & $8.69\times10^{-7}$ & 0.9654 & 0.0035 \\ 
10 & 55 & 2.805 & 0.400 & $5.62\times10^{-6}$ & $8.01\times10^{-7}$ & 0.9695 & 0.0030 \\ 
10 & 60 & 2.837 & 0.400 & $5.29\times10^{-6}$ & $7.45\times10^{-7}$ & 0.9727 & 0.0027 \\ 
\bottomrule
\end{tabular}
\end{table}

Because Eq.~\eqref{Nelf} is derived within the slow-roll expansion, it should not be interpreted as an exact background integration all the way through the exit.  At horizon crossing the parameters entering the observables are small in the displayed range, but near $\epsilon_1=1$ some higher parameters can become of order unity.  For the scan in Eq.~\eqref{eq:numericalscan}, the conditions $\Delta_1\leq1$ and $\epsilon_2\leq1$ are recovered roughly within the final half e-fold before the end, while $|k_2|\leq1$ is recovered a few e-folds before the end, depending on $m_1$.  This behaviour is known in Gauss-Bonnet inflationary models \cite{Pozdeeva:2021nmz}.  The numerical curves are therefore leading slow-roll estimates; a precision determination of the last e-folds and of the reheating matching would require solving the full background system, Eqs.~\eqref{e0025}--\eqref{e0027}, without imposing the slow-roll truncation.

\begin{figure}[H]
\centering
\includegraphics[width=0.95\textwidth]{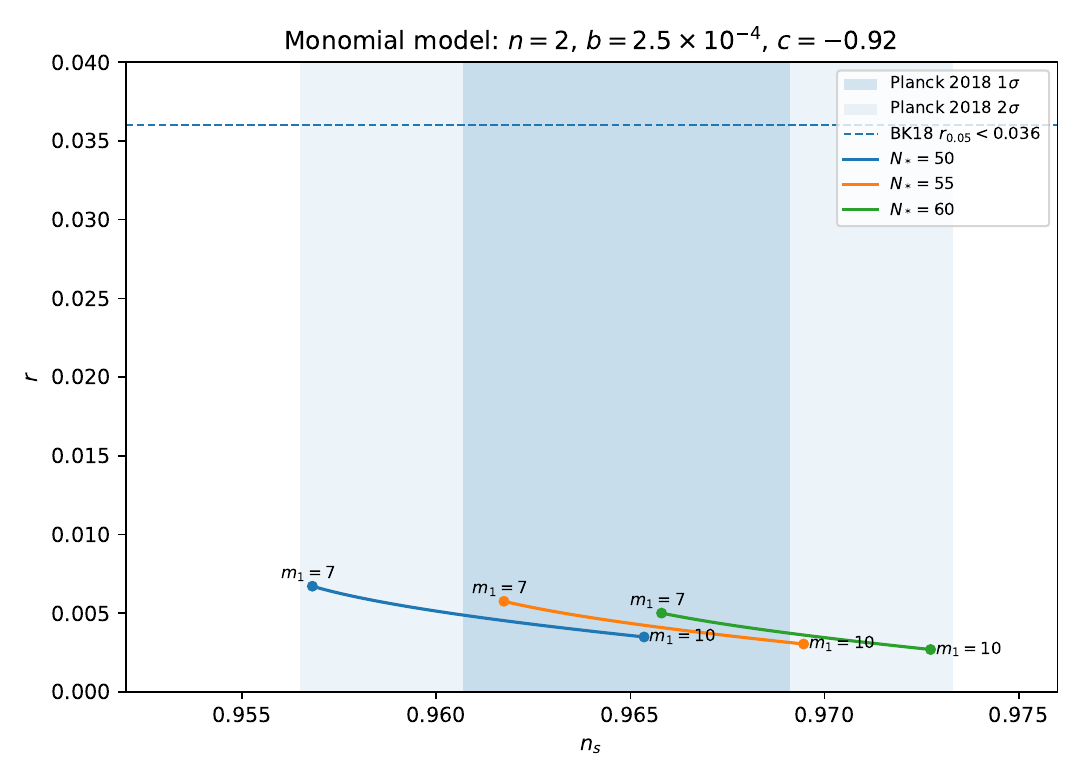}
\caption{Scalar spectral index $n_S$ versus tensor-to-scalar ratio $r$ for $N_*=50,55,60$ in the representative scan $n=2$, $b=2.5\times10^{-4}$, $c=-0.92$, and $7\leq m_1\leq10$.  The shaded bands show the Planck 2018 one- and two-standard-deviation intervals for $n_S$, and the dashed horizontal line shows the BK18 bound $r_{0.05}<0.036$.}
\label{GBns_r}
\end{figure}

\begin{figure}[H]
\centering
\begin{tabular}{cc}
\includegraphics[width=0.48\textwidth]{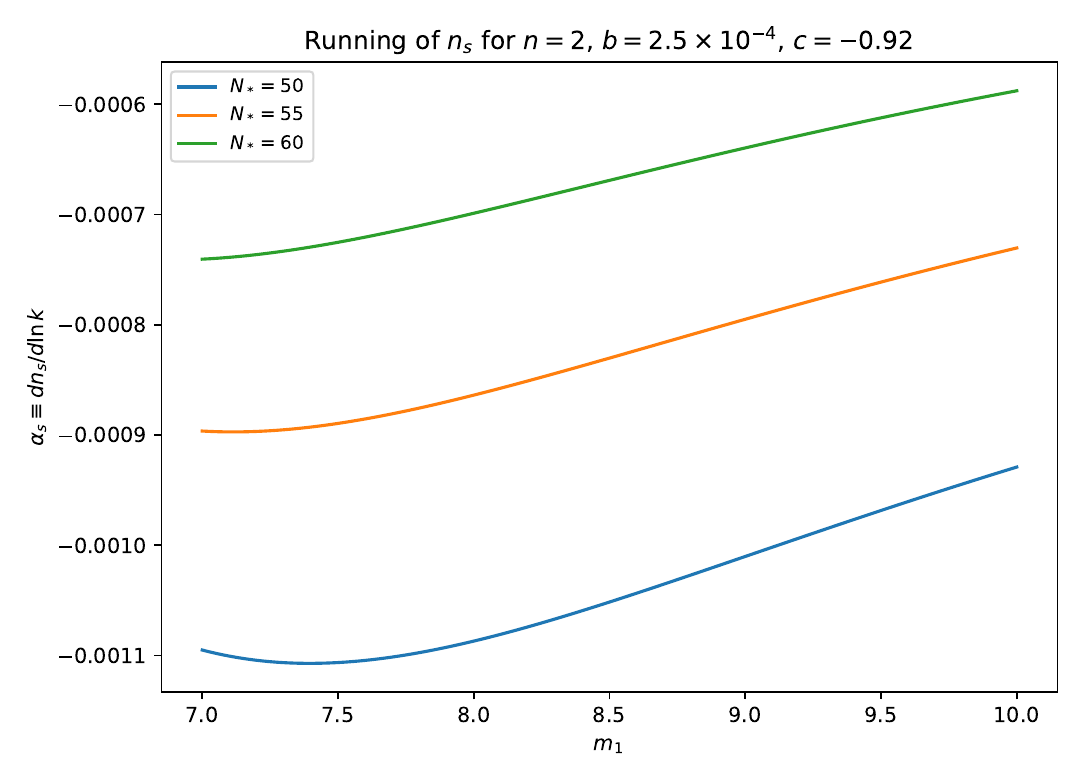} &
\includegraphics[width=0.48\textwidth]{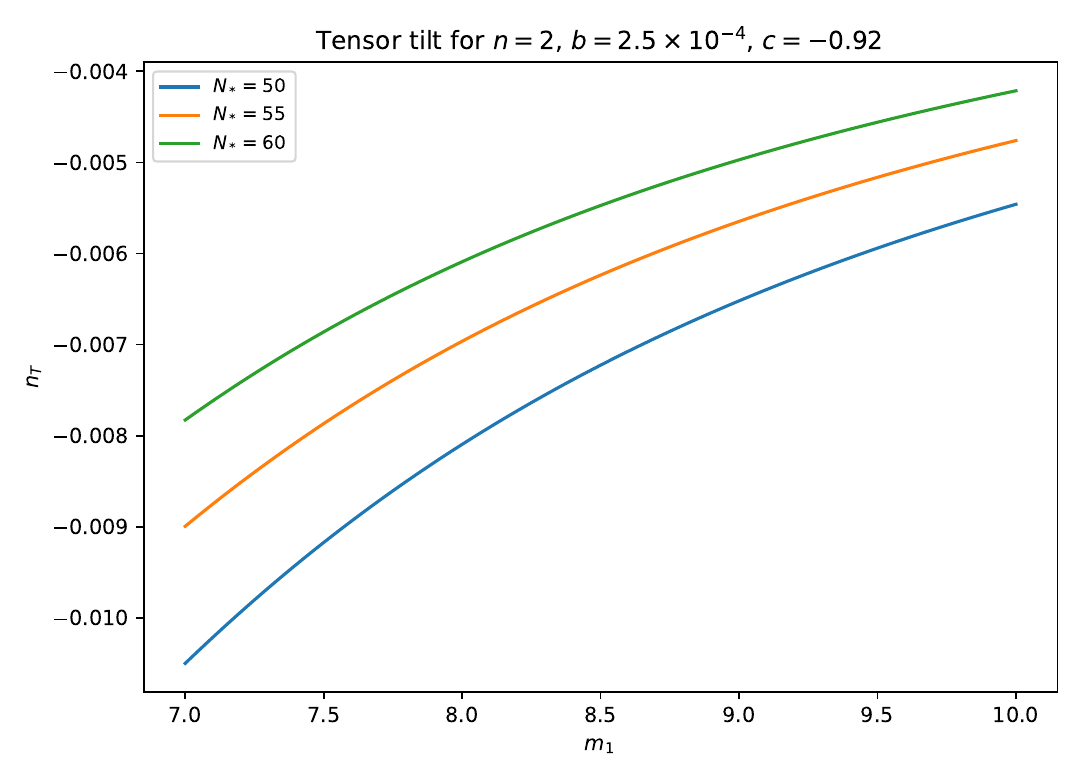}
\end{tabular}
\caption{Running of the scalar spectral index $\alpha_s\equiv dn_S/d\ln k$ (left) and tensor spectral index $n_T$ (right) as functions of $m_1$, for $n=2$, $7\leq m_1\leq10$, $b=2.5\times10^{-4}$, and $c=-0.92$.}
\label{GBdns_m}
\end{figure}

For illustration, we present representative numerical results for this toy model.  Figure~\ref{GBns_r} shows the correlation between the tensor-to-scalar ratio $r$ and the scalar spectral index $n_S$ for the scan in Eq.~\eqref{eq:numericalscan}.  The endpoint with larger $r$ corresponds to $m_1=7$, while the lower-$r$ endpoint corresponds to $m_1=10$.
Figure~\ref{GBdns_m} shows the running of the scalar spectral index and the tensor spectral index for the same parameter range.


\section{Conclusions and further remarks}
\label{sec:5}

We have investigated the use of CMB observations to constrain scalar-tensor models of gravity beyond GR.  We focused on slow-roll inflation in models with minimal coupling to the Ricci scalar, non-minimal derivative coupling to the Einstein tensor, and scalar coupling to the Gauss-Bonnet invariant.  After deriving the general slow-roll equations, we analysed monomial choices for the potential, the kinetic coupling, and the Gauss-Bonnet coupling.

At leading slow-roll order, the inflationary predictions depend on a small set of combinations of the functions appearing in the action.  We derived explicit expressions for $n_S$, $n_T$, $r$, and their running quantities, and we displayed consistency relations that quantify the departure from the standard single-field relation $r=-8n_T$.  In the present class of models, the Gauss-Bonnet contribution modifies that relation through the parameter $\Delta_1$.

The monomial example illustrates how non-minimal kinetic and Gauss-Bonnet interactions can lower the tensor-to-scalar ratio relative to minimally coupled monomial potentials.  This is important because the final Planck 2018 constraints and the BK18 BICEP/Keck bound strongly disfavor minimally coupled monomial potentials with $n\geq2$ over much of the standard parameter range \cite{planck18,BK18}.  The representative numerical curves shown above should therefore be read with the current tensor bound in mind: the phenomenologically viable region is concentrated at lower values of $r$.  We have also made explicit that the e-fold number computed from the slow-roll integral is an approximation; near the end of inflation higher slow-roll parameters can become order unity, and a full background integration is required for a precision exit analysis.

Future CMB polarization measurements will further test this class of scenarios by improving the sensitivity to primordial tensor modes and to the scale dependence of the scalar spectrum \cite{BICEPKeck2024}.  The formulae collected here provide a useful starting point for translating such data into constraints on the scalar-tensor functions $V(\varphi)$, $\ffuno(\varphi)$, and $\ffdos(\varphi)$.

\section*{Acknowledgments}
The work of ETL has been supported in part by the Ministerio de Educaci\'on y Ciencia, grants FIS2025-24924,
 Universidad de Murcia project E024-018 and Fundacion Seneca (21257/PI/24).
ETL acknowledges useful and motivating conversations with L. Granda during the early stages of this work.


\end{document}